\documentstyle[preprint,aps]{revtex}
\begin{document}
\draft
\tightenlines
\title{
\bf{ POLYMERS WITH RANDOMNESS: PHASES AND PHASE TRANSITIONS
} }
\author{Somendra M. Bhattacharjee\cite{eml} and Sutapa
Mukherji\cite{eml1} \\ }
\address{ Institute of Physics,
Bhubaneswar 751 005, India}
\maketitle
\begin{abstract}
We discuss various aspects of the randomly interacting directed
polymers with emphasis on the phases and phase transition. We
also discuss the behaviour of  overlaps of directed paths in a
random medium.
\end{abstract}

\pacs{ }

\newpage

\section{Introduction}
Polymers with randomness can be classified into two categories:
(1) random medium, and (2) random interaction, analogous to the
random field and random bond problems for magnets.
These random problems remain notoriously difficult \cite{bkc}.
Attention has therefore
gradually shifted to simpler models, and for the past ten years
directed polymers played a crucial role in unraveling various
issues concerning disordered systems.  This model is also
important because of its rich behavior,
and connection with nonequilibrium surface growth problem with
noise, flux line pinning in dirty samples etc \cite{tim}.
We like to give an overview of the problem of randomly
interacting directed polymers.

The pure problem is discussed in the second section.  The
randomly interacting model and its field theoretic study can be
found in section III. An exact real space renormalization group
approach to study the finite size effect is given in section IV.
The question of overlap of two paths in a random medium that can
be recast as a problem of interacting directed polymers in a random
medium is explored in section V.  A summary is given in section
VI.

\section{Pure problem: binding-unbinding}
Directed polymers (DP) are lines or polymers with a preferred
direction. For $m$ such DPs
a general Hamiltonian with many body interaction can be written as:
\begin{equation}
H_m= \frac{1}{2}\sum\limits_{i=1}^m\int\limits_0^N dz
\ {\dot {\bf r}}_i^2 + \int\limits_0^N dz \ v_m \ \prod_{i=1}^{m-1}
\delta^d ({\bf r}_{ii+1}(z)) +
\sum\limits_{i=1}^m\int\limits_0^N dz \ V({\bf r}_i,z),
\label{eq:hn}
\end{equation}
where ${\dot {\bf r}}_i=\partial {{\bf r}}_i/\partial z$, and
${\bf{r}}_{ij}(z) = {\bf r}_{i}(z)- {\bf r}_{j}(z),\
{\bf r}_i (z) $ being the $d$ dimensional transverse coordinate of a
point at $z$ on the contour of the $i$th chain
\cite{jja,jjb,smbphy,lassl}.  The first
term is the elastic energy, taking care of the connectivity
of the chains.  The polymers interact with a coupling constant
$v_m$ if all the $m$ chains meet at a point.  There can also be
an external potential $V({\bf r},z)$ which in the random
potential problem is a stochastic variable. For most of this
paper we will consider only $m=2$ and $m=3$ cases.  The external
random potential problem will be considered in the last section.

With $V=0$, the polymers undergo a binding unbinding transition as the
strength of the potential is varied. For the two body ($m=2$) problem,
the transition takes place at $v_2=0$ for $d<2$, while a minimum
strength of attraction is needed for binding at $d>2$.  This is reflected
in the renormalization group (RG) approach through the unstable fixed
point for the $\beta$ function for the coupling constant.  This
$\beta$ function tells us the flow of the coupling as the system
is probed at a bigger length scale. It is now well known that
the field theoretic RG can be implemented exactly for this
class of many body problems.  Introducing a dimensionless
coupling constant $u_2 = v_2 L^{2-d}$, the exact $\beta$
function is given by
\begin{equation}
\beta(u)\equiv L\frac{\partial u}{\partial L}= 2 \epsilon' u
\left ( 1-\frac{u}{4\pi \epsilon'}\right )\label{eq:bu},
\end{equation}
where $u$ is the renormalized dimensionless coupling constant
\cite{jja}.  Note that $2{\epsilon}'=(2-d)$.

The flow diagram for the dimensionless coupling constant
$u$ is shown in Fig.1a.  The fact that for $d<2$
any small attractive interaction is able to form a bound
state is reflected by the flow to the nonperturbative
regime for any negative $u$. The repulsive or the positive
$u$ region is dominated by the stable, ``fermionic" fixed point
$u^*(=4\pi{\epsilon}')$. For $d>2$ there exists a
nontrivial unstable fixed point $u=u^*$ which separates the
bound and the unbound states for the two polymers. In
short, the unstable fixed point represents the critical
point for the binding-unbinding transition. The correlation
length, $\xi_{\parallel}$ for the transition describes the
average separation
of two contacts along the chain, and it diverges as the critical
point is approached with an exponent
$\nu=1/{\mid{\epsilon}'\mid}$ for $1\leq d< 4$, except for
$d=2$, where the correlation length diverges exponentially as
$\exp (1/v_2)$.  Other
approaches seem to suggest that $d=4$ is the upper critical
dimension for this problem \cite{lip,majum}, however RG is yet
to give us that result.  We would like to point out that the
exact $\beta$ function of Ed. \ref{eq:bu} is obtained by
absorbing the poles at $d=2$ in a dimensional regularization
scheme.  This, of course, leaves the poles at $d=4$ untreated.
Is it the signal for an upper critical dimension at $d=4$?

The stable fixed point describes, in this problem, the
unbound phase.  A way of characterizing the phase is to look at
the asymptotic behaviour of the reunion partition function,
${\cal Z}_R(N)$.  This partition function describes the
situation where the chains are tied together at both the ends,
and the end  points can be anywhere in space.
The asymptotic behaviour of ${\cal Z}_R(N) \sim N^{-\Psi}$ was
studied long ago for $d=1$ in a different context \cite{fish84,hf}.
It is known that $\Psi=3/2$ for two chains in 1 dimension with
repulsive two body interaction.  It follows
from an exact random walk analysis or from a use of fermionic nature
of the quantum particles.  These methods are restricted to one
dimension only.  RG is the appropriate framework to obtain the
asymptotic behaviour for other $d$, and the exponent follows
form the multiplicative renormalization constant for the
partition function.  In this framework, the Huse-Fisher 1
dimensional exponent can be recovered
as an $O(\epsilon')$ result which is, in fact,
exact\cite{reujphys,reupre}. Furthermore,
this RG analysis also shows that because of the marginality of the
coupling at $d=2$, there is a special log correction to the
Gaussian behaviour, and ${\cal Z}_R(N) \sim N^{-1} (\log
N)^{-2}$. More general results can be found in Ref.
\cite{reupre}. This log correction in two dimensions has recently
been recovered by Guttmann and Essam in an exact lattice
calculation\cite{gutt}.

If we now go to the three body interaction, then again a similar
exact analysis can be carried out \cite{jjb}.  We, in this
paper, however,
restrict ourselves only to $d=1$ which turns out to be the
marginal case for $v_3$.  The critical exponent for unbinding
transition is $\nu=2/\mid 1-d\mid$, except for $d=1$ where
the correlation length diverges like $\exp (1/v_3)$.  The
three chain reunion partition function will have a log
correction, identical to the marginal two chain case, namely,
${\cal Z}_R(N) \sim N^{-2} (\log N)^{-2}$.  The similarity in
the behaviour of the many-body interactions, if proper variables
are used, is a novel feature of the multicritical directed
polymers, and has been termed ``Grand Universality" \cite{smbphy}.
We will see that such a {\it grand universality} is preserved also
for the random case.

Since there is only one fixed point at $d=2 (d=1)$ for the two
(three) chain problem, one can identify
the fixed point at zero as the critical point while the approach
to the fixed point as describing the phase of the system.

\section{Random interaction}
We now consider the random version where the polymers interact
with a random coupling constant and there is no external
potential. For simplicity we consider
randomness to be dependent only on $z$.  A physical realization
would be a random distribution of monomers (or charges) along
the backbone.  The interaction is given by
\begin{equation}
\int_{0}^{N} dz \ v_0\ [1+b(z)]\ \delta{\bf
(}{\bf{r}}_{12}(z){\bf )},
\label{eq:h}
\end{equation}
where the randomness is introduced through $b(z)$.  We take
uncorrelated disorder with a Gaussian distribution
\begin{mathletters}
\begin{eqnarray}
P{\bf (}b(z){\bf )}=(2\pi\Delta)^{-1/2}\
\exp[-b(z)^2/(2\Delta)],
\label{eq:disa} \\
\langle b(z)\rangle=0, \ {\rm and}\ \langle b(z_1)
b(z_2)\rangle\ =\Delta\
\delta(z_1-z_2).\label{eq:disb}
\end{eqnarray}
\end{mathletters}

For the many body interaction problem, the random Hamiltonian
would be similar to Eq. \ref{eq:hn} except that the coupling
constant $v_m$ is to be replaced by $v_m(1+b(z))$ inside the
integral.

The approach we take is to study the various cumulants of the
partition function.  The first cumulant, as one might expect,
behaves like the pure problem but with a shifted critical point.
Since there are sites with attractive interaction, the critical
point for unbinding occurs not at zero average interaction but
at a certain nonzero repulsion.  It would also mean that even if
the chains are on the average repulsive, (i.e., $v_2 >0$), a
binding-unbinding transition can be induced by tuning the
disorder or changing the temperature. Such a thermal unbinding
is not possible in the pure case for $d \leq 2$.  Apart from
that, the critical behaviour remains the same.  The situation is
different for the second cumulant.

When we consider the second cumulant of the partition function,
we require four (six)
chains for the two (three) body case.  On averaging over the
disorder, an interaction (``inter replica" interaction) is
generated that couples the original chains with the replica.
For example, for the two body problem, the interaction is of the
type
\begin{equation}
H_{\rm rep}=-\bar{r}_0\int_0^N dz\ \delta{\bf (}{\bf r}_{12}(z){\bf
)}\
\delta{\bf (}{\bf r}_{34}(z){\bf )},\label{eq:zse}
\end{equation}
with ${\bar{r}}_0=v_0^2\Delta$.  This interaction, different
from the four body multicritical interaction of Eq. \ref{eq:hn},
is a correlation effect. If chains 1 and 2 meet at length $z$,
then it is favourable for 3 and 4 also to have a contact at that
same $z$ though not necessarily at the same transverse space
coordinate.  The importance of the disorder can therefore be
understood if we know the flow of this interaction as the
probing length scale is changed.  If we are at the critical
point of the average interaction, then RG can be implemented
exactly \cite{smprl,smpre}.  Defining the dimensionless coupling
constant $r_0$ through an arbitrary length scale $L$ as
$r_0=\bar{r}_0L^{2\epsilon}(4\pi)^{-d}$, $\epsilon=1-d$, and
denoting the renormalized coupling as $r$, the RG $\beta$
function is given by
\begin{equation}
\beta(r)\equiv L\frac{\partial r}{\partial L}=2 (\epsilon
r+r^2)\label{eq:br}.
\end{equation}
There are two fixed points: (i) $r=0$ and (ii)
$r^*=-\epsilon$. See Fig. 1b.
The bare coupling constant $r_0$ which originates from
$v_0^2\Delta$, where $\Delta$, the variance of the
distribution, is strictly positive, requires a positive
$r$. Therefore, the nontrivial fixed point for $d<1$ in
negative $r$ is unphysical.  It however moves to the
physical domain for $d>1$.
The main feature
that comes from the analysis is that the disorder is marginally
relevant at $d=1$. This means that even a small disorder will
change the nature of the unbinding transition and take the
critical system to a ``strong" coupling regime.  There is
however no perturbative fixed point to describe the strong
coupling regime.  A marginally relevant variable means that a
new critical feature would appear for higher dimensions.  This
is reflected in the new unstable fixed point.  For small enough
disorder, the $\beta$ function for $d>1$ takes $\Delta$ to
zero, reproducing a pure type behaviour.  If however, the
starting disorder is higher than the fixed point value, then it
goes to the strong coupling regime.  The unbinding transition is
therefore pure type for small disorder (``weak" disorder) and
beyond a critical disorder, in the ``strong" disorder regime,
a new critical behaviour is expected.

The exact nature of the RG is lost if $v_2 \neq 0$.  In a one
loop approach, there are indications of the existence of a fixed
point for the stable fixed point of $v_2$ \cite{smrg}.  Since
the flows are
different on the two sides, one wonders whether this indicates a
new phase also.

The exact $\beta$ function of Eq. \ref{eq:br} tells us also that
around the critical disorder, the relevant length scale exponent
is $1/\mid 1 - d\mid$ along the chain.  In one dimension, the
length scale diverges exponentially around $r=0$.  These exponents
have been verified numerically.

A dynamic renormalization group approach has also been developed
for the two chain problem.  In this approach, instead of looking
at the moments of the partition function, the free energy is
probed directly.  This approach yields the same results and
shows that there is no special fluctuation exponent for the free
energy.  \cite{kala}

\section{Real space RG}

Due to the absence of any fixed point for the strong coupling
regime, it is necessary to study the problem using
nonperturbative methods.  One such method is the real space RG
(RSRG), which can be implemented
exactly on hierarchical lattices\cite{derr,scar} as shown in Fig
2.  To avoid
unnecessary approximations, we work with these lattices from the
beginning.  As per construction, we consider the partition
function for two chains tied at the both ends and with a random
site interaction.  We want to study the various moments of the
partition function.

It is again clear that for the $n$th moment, we
require $2n$ chains and they will be coupled by the disorder.
An effect of this is that there is an analog of the
binding-unbinding transition for each moment, and the higher the
moment ($n$)
the higher is the transition temperature.  In the high
temperature phase for the $n$th moment, the free energy is
expected to approach the free entropy of $2n$ chains.
Subtracting out this free part, the free energy approaches zero
in the thermodynamic limit for high temperatures while, it has
an $O(1)$ value per bond in the low temperature phase.  Let us
define ${\cal Z}_{\mu} (n)= {\overline Z_{\mu}^n}/S_{\mu}^{2n}$,
where $\mu$ is the generation, $S_{\mu}$ is the entropy of a single
chain of length $L_{\mu} = 2^{\mu-1}$, the overline representing
the disorder average.  If we
keep the temperature fixed (above the critical point for the
first moment), then there exists a critical $n$, $n_c$, so that
for $n <n_c$ the moments are in the high temperature phase.    We want
to study the finite size correction to the moments of as
$n\rightarrow n_c-$.

 The approach to the thermodynamic limit can be written
as
\begin{equation}
{\cal Z}_{\mu}(n) = {\cal Z}^{*}(n) + {B_z(n)}\ {L_{\mu}^{-\psi}}
+ ...,
\label{eq:geng}
\end{equation}
where ${\cal Z}^{*}(n)$ is the thermodynamic limit ($\mu \rightarrow
\infty$), and $B_z(n)$
is the amplitude of the finite size correction.

For a given realization of the disorder, the partition function can
be written as (see Fig 2)
 \begin{equation}
Z_{\mu+1}= b Z_{\mu}^{({\cal A})} y Z_{\mu}^{({\cal B})} +
b (b-1) S_{\mu}^2. \label{eq:zrec}
\end{equation}
The first term originates from the configurations that require
the two DPs to meet at $C$, while the second term counts the "no
encounter" cases. There are no energy costs at the two end
points. The Boltzmann weight is random, and for a Gaussian
distribution of energy,  ${\overline{ y^m}} =
{\overline{y}}^{m^2}$.  Note also that $S_{\mu} = b^{L_{\mu}
-1}$, where
$L_{\mu}=2^{\mu}$ is the length of DP.

The moments of the partition function, from Eq. \ref{eq:zrec},
are
\begin{equation}
{\cal Z}_{\mu+1}(n) = b^{-n} \sum_{m=0}^{n} P_{nm} {\cal
Z}_{\mu}^2(m) ,
\label{eq:rec1}
\end{equation}
where $P_{nm}={n \choose m} (b-1)^{n-m}{\overline{y^m}}$, with
the initial condition ${\cal Z}_0(n) = 1$ for all the moments
because there is no interaction in the zeroth generation (one
single bond).
By iterating these recursion relations, the moments are
calculated exactly and the finite size correction is estimated.
As shown in Fig. 2, the amplitude has a power law divergence
as $n_c$ is approached, namely
\begin{equation}
B_z(n) \sim (n_c-n)^{-r}, \quad {\rm for} \ n\rightarrow
n_c-,\label{eq:bexr}
\end{equation}
with $r=.71\pm .02$.  The exponent is independent of temperature
but depends on the effective dimensionality of the system. We
call this a scar left by disorder in the high temperature phase.

\section{Overlap in a random medium}

Much has been achieved in  the problem of directed
polymer in a random medium (DPRM). Unlike the random interaction case,
carrying out a systematic RG beyond one loop is extremely
hard \cite{frey}.
On the other hand a remarkable extra gain in the DPRM case is the exact
knowledge of the nontrivial
geometric and  thermal properties at least at
$d=1$. It is known through the mapping of DPRM problem to the
nonlinear noisy growth equation of a surface (KPZ equation)
\cite{kpz} that at
$d=1$ the
transverse extension of the polymer and the free energy
fluctuation scale as ${\overline{<x>^2}}^{1/2}\sim N^{\zeta=2/3}$ and
$f\sim x^{\chi=1/2}$.
These results can be proved to be exact through the fluctuation
dissipation theorem. At $d=1$ this new value of $\zeta \neq 1/2$
persists for all temperatures and the system is always at the
strong disorder or ``low temperature" phase. For $d>2$ there is
a transition from a
high temperature phase (free polymer) to a low temperature (
`spin glass' type) phase. Though the unstable fixed point is
well under control, the strong disorder fixed point is
not reachable through perturbation. Numerical approaches
intensified the controversies about the strong disorder phase.
Another unsolved question is the
existence of an upper critical dimension (UCD) which, in some
approaches, seems to be 4.  There is a hope that if RG can
resolve the question of UCD for the pure interacting DP problem,
as mentioned in section II, then the UCD problem for DPRM can
also understood.

Here we discuss how the overlap behaves in this
problem. Since the low temperature phase is a spin glass type
phase we expect the overlap to serve as an order parameter. The m
chain overlap is in general defined as
\begin{equation}
q_m=-\frac{1}{t}\int_0^t dz \ {\overline{<\prod _{i=1}^{m-1} \delta
({\bf x}_{i,i+1}(z))}} >,
\end{equation}
where ${\bf x}_{i,i+1}(z)={\bf x}_i(z)-{\bf x}_{i+1}(z)$, bar
and the angular brackets denote the disorder and  thermal
average respectively. In the replica language, this overlap
describes the common configurations of the valleys (i.e. common
to $m$) of a rugged free energy landscape.  The overlap comes
from the common path of $m$ chains in the same random medium.
The behaviour of the overlap is nontrivial because the disorder
induces an attraction among the replicas.
We consider only the two chain overlap here.

The main key to solve the problem is to introduce a $2$ body
interaction in the Hamiltonian of DPRM problem and
use a mapping that leads to a KPZ type nonlinear equation \cite{smrap}.
Defining the quenched free energy as $f_2(v_2,z)$, the
overlap can be obtained as $q_2=-z^{-1}\
df_2(v_2,z)/dv_2\mid_{v_2=0}$. With the scaling hypothesis
for the free energy, $f_2=z^{\chi/\zeta}{\sf
f}(v_2t^{-\phi\zeta})$ we find the scaling of the overlap
as $q_2\sim z^{\Sigma}$ with $\Sigma=(\chi-\phi-1)\zeta$. Our
interest here is in finding out $\Sigma$.

The working Hamiltonian is therefore similar to Eq.
\ref{eq:hn} with $m=2$.  For convenience, we introduce a line
tension $\gamma$ so that the elastic part of Eq. \ref{eq:hn} is of
the type $\frac{\gamma {\dot {\bf r}}_i^2}{2}$ and also choose the
coupling constant as $\lambda v_2/(2\gamma)$ and $\lambda V/(2\gamma)$
as the random potential.  The random potential is with zero mean and
$\langle V({\bf r},\tau) V({\bf r}',\tau')\rangle =
\Delta \delta({\bf r}-{\bf r}')\delta(\tau-\tau')$.
The free energy
defined as $h(\{{\bf x_j}\},t)=(2 \gamma/\lambda) \ln Z(\{{\bf
x_j}\},t)$, where $Z(\{{\bf x_j}\},t)$ is the partition function,
 satisfies
\begin{equation}
\frac{\partial h}{\partial  t}=\sum\limits_{j=1}^{2}(\gamma
\nabla^2_jh+\frac{\lambda}{2}(\nabla_j h)^2)+g_0,
\end{equation}
where $g_0=\sum_{j=1}^{2}V({\bf
x}_j,t)+v_2\delta({\bf x}_{12}(t))$.
In order to bring out the similarity with growth equation we
use $t$ instead of $z$ as the variable for the special direction.
A dynamic renormalization group calculation upto $O(\lambda^2)$
and $O(v_2)$ for the above equation in the Fourier space leads
to the following crucial results\cite{smrap}.
({\bf i.}) There is no change in the single chain behavior. The
renormalization of the single chain propagator, and the
nonlinearity is the same as that of a single chain in a random
medium.
({\bf ii.}) The coupling constant gets renormalized even at
$O(v_2)$ as given below
\begin{equation}
v_{2R}=v_2+8(-\frac{\lambda}{2})^2(2v_2\Delta)\int_{p,\Omega}
p^4G_0({\bf p},\Omega)[G_0(-{\bf p},-\Omega)]^2G_0({\bf p},-{\bf
p},0).
\end{equation}
Here $G_0$ is the bare propagator, defined as $G_0(\{{\bf
k}_j\},\omega)=[\gamma \sum_j k_j^2-i\omega]^{-1}$, $({\bf
k},\omega)$ being the Fourier conjugate variables for $({\bf
r},t)$. A short hand notation $\int_{{\bf p}\omega} \sim \int
d{\bf p} d \omega /(2\pi)^{d+1}$ is used. In the
above equation only the nonzero momentum vectors are written
explicitly as the arguments of $G_0$.

The next obvious point is the presence of an anomalous dimension
in the coupling constant as is apparent from the recursion
relation
\begin{equation}
\frac{dv_2}{dl}=[z-\chi-d+U]v_2.
\end{equation}
Here the term proportional to
$U=K_d\lambda^2\Delta/(2\gamma^3)$ is the anomalous part from the
RG and $K_d=(2\pi)^{-d} S_d$, with $S_d$ being the surface
area of the unit $d$ dimensional sphere.
Since we find that the single chain properties remain
unaffected, we use in the following the KPZ fixed point value
for $U$. Let us first consider $d=1$.
It has been shown that the
KPZ fixed point is a stable one and indicates a glassy behavior
at all temperatures. By substitution $U^*=1$, we find that
$\Sigma=0$, which one would expect in the low temperature phase,
as one finds numerically \cite{mez}.
For $d=2+\epsilon$ the KPZ fixed point
$U^*=2\epsilon$, being an unstable fixed point,
corresponds to the spin glass transition.
The exponent for the overlap can be readily obtained
at this fixed point as $\Sigma=-[d+\eta]$, where
$\eta=-2\epsilon$. The analysis can be extended to $m$ chain
overlap also.

\section{Conclusion}

Randomly interacting directed polymers exhibit a weak to strong
disorder transition for $d>1$.  This can be established by an
exact renormalization group approach.  Real space
renormalization group approach for hierarchical lattices reveals
a diverging finite size correction.  This might indicate the
existence of a phase with no counterpart in the pure system.
Could it be a Griffiths phase? This is an open question.
Using interacting directed polymers in a random medium, we have
calculated the decay of overlap at the spin glass transition
point in $2+\epsilon$ dimensions. In one dimension, this
analysis recovers the result based on fluctuation arguments and
numerical simulation.

\begin{figure}
\caption
{Flow diagrams for (a) the pure coupling $u$ for two chains, and
(b) the disorder $r_m$ for $m$ chains. Here $d_m = 1/(m-1)$.
In both cases, * denotes nontrivial fixed points. }\label{flo}
\end{figure}
\begin{figure}
{Plot of $B_z(n)^{-1/r}$ vs $n$ for $b=4$ and various  temperatures.
(a) $\log
{\overline{y}} = 0.065$, and $r=0.73$, (b) $\log {\overline{y}}
= 0.04$, and $r=.73$ (c) $\log {\overline{y}} = 0.03$, and
$r=0.72$ (d) $\log {\overline{y}} = 0.02$, and $r=0.72$. Inset
shows the construction of a hierarchical lattice with $b=2$.
.}\label{hier}
\end{figure}

\begin{references}
\bibitem[*]{eml1} electronic address: sutapa@iopb.ernet.in
\bibitem[**]{eml} electronic address: sb@iopb.ernet.in
\bibitem{bkc} K. Barat and B. K. Chakrabarti, Phys. Rept., {\bf
258}, 377 (1995).
\bibitem{tim} For a recent review, see, e.g., T. Halpin-Healy
and Y. C. Zhang, Phys. Rept., {\bf 254}, 215 (1995).
\bibitem{jja} J. J. Rajasekaran and S. M. Bhattacharjee,
J. Phys.  {\bf A24}, L371 (1991).
\bibitem{jjb}S. M. Bhattacharjee and J. J. Rajasekaran, Phys.
Rev. A {\bf 46}, R703  (1992).
\bibitem{smbphy} S. M. Bhattacharjee, Physica A{\bf 186}, 183
(1992).
\bibitem{lassl} M. L\"assig, Phys. Rev. Lett. {\bf 73}, 561
(1994); {\bf 74}, 3089(1995) (E).
\bibitem{lip} R.Lipowsky Europhys. Lett. {\bf 15},
703(1991)
\bibitem{majum} S. N. Majumdar Physica A {\bf 169}, 207 (1990).
\bibitem{fish84} M. E. Fisher, J. Stat. Phys. {\bf 34},
667 (1984).
\bibitem{hf} D. A. Huse and M. E. Fisher, Phys. Rev.
{\bf B29}, 239 (1984).
\bibitem{reujphys} S. Mukherji and S. M. Bhattacharjee, J. Phys.
A{\bf 26}, L1139 (1993); erratum (in press).
\bibitem{reupre} S. Mukherji and S. M. Bhattacharjee, Phys. Rev.
E {\bf 48}, 3427 (1993); erratum (in press).
\bibitem{gutt} J. Essam and A. J. Guttmann, private communication
\bibitem{smprl} S. M. Bhattacharjee and S. Mukherji, Phys. Rev.
Lett {\bf 70}, 49 (1993).
\bibitem{smpre} S. Mukherji and S. M. Bhattacharjee, Phys. Rev. E
{\bf E48}, 3483 (1993).
\bibitem{smrg} S. Mukhherji, unpublished.
\bibitem{kala} H. Kallabis and M. L\"assig, Phys. Rev. Lett. {\bf
75}, 1578 (1995).
\bibitem{derr} J. Cook and B. Derrida J. Stat. Phys {\bf 57}, 89
(1989); E. Medina and M. Kardar ibid {\bf 71} 967 (1993).
\bibitem{scar} S. M. Bhattacharjee and S. Mukherji, Phys. Rev.
E (in press).
\bibitem{frey} E. Frey and C. T\"auber, Phys. Rev. E {\bf 49}
(1994).
\bibitem{kpz} M. Kardar, G. Parisi and  Y. C. Zhang, Phys. Rev.
Lett. {\bf 56}, 889 (1986).  M. Kardar and Y. C. Zhang,
Phys.  Rev. Lett. {\bf 58}, 2087 (1987).



\bibitem{smrap} S. Mukherji Phys. Rev. {\bf E50} R2407 (1994).
\bibitem{mez} M. Mezard, J. Phys.(Paris) {\bf 51}, 1831 (1990).

\end{references}
\end{document}